\documentclass[twoside]{article}
\usepackage{latexsym,amsmath,amssymb}
\usepackage{cite} 
\usepackage{sectsty}
  \sectionfont{\large} \subsectionfont{\normalsize}
\usepackage{graphicx}
\usepackage{theorem}
\usepackage[all]{xy}

{\theorembodyfont\rmfamily

\newtheorem{corollary}{Corollary}

}
\newenvironment{proof}{\begin{paragraph}
          {Proof}}{\end{paragraph}}

\renewenvironment{abstract}
 {\small\begin{quote}{\textbf{Abstract}}\,\,}{\end{quote}}
\newenvironment{keywords}
 {\small\begin{quote}{\textbf{Keywords}}\,\,}{\end{quote}}
\newenvironment{classification}
 {\small\begin{quote}{\textbf{2010 Mathematics Subject Classification}}\,\,}{\end{quote}}
\date{}

\title{\vspace{-9ex}
{\centering
 \textbf{\large A Method of Reducing Dimension of Space Variables in Multi-dimensional Black-Scholes Equations}}}

 \vspace{-4ex}

\author{\small\textsf{\bfseries 
$^{1}$ Hyong-chol O, $^{2}$Yong-hwa Ro $^{3}$Ning Wan}\\[-.5ex]
{\footnotesize  ${}^{1, 2}$ Faculty of Mathematics, \textbf{Kim Il Sung} University}\\[-1ex]
{\footnotesize   Pyongyang , D. P. R. of Korea}\\[-.5ex]
{\footnotesize  ${}^3$Department of Applied Mathematics, Tong-ji University}\\[-.5ex]
{\footnotesize  Shanghai, China}\\[-.5ex]
{\footnotesize e-mail:  $^{1}$ohyongchol@yahoo.com, $^{2}$leewunghun@yahoo.com,
$^{3}$blumping@etang.com}}

\begin{document}

\maketitle
\thispagestyle{empty}

\vspace{-.6cm}

\begin{abstract}
We  study  a method of reducing space dimension in multi-dimensional Black-Scholes partial differential equations as well as in multi-dimensional parabolic equations. We prove that a multiplicative transformation of space variables in the Black-Scholes partial differential equation reserves the form of Black-Scholes partial differential equation and reduces the space dimension. We show that this transformation can reduce the number of sources of risks by two or more in some cases by giving remarks and several examples of financial pricing problems. We also present that the invariance of the form of Black-Scholes equations is based on the invariance of the form of parabolic equation under a change of variables with the linear combination of variables. 
\end{abstract}



%
%

\section{Introduction}

Multi-assets option prices satisfy multi-dimensional Black-Scholes partial differential equations \cite{jia}.
It is well known that the change of numeraire gives very important computational simplification in multi-assets option pricing. See \cite{ben}.  But using the technique of numeraire change, we can reduce the number of sources of risks by one. See  \cite{hyo}.  

On  the  other  hand,  in  some  pricing  problems  of  financial  derivatives,  we  can  see  some transformations of variables reducing the number of sources of risks by two or more. For example, in \cite{xug}, they reduced the number of sources of risks from 3 to 1 in pricing a European call foreign currency option. Their main method is to combine the change of numeraire and the transformation of multiplication of two variables. Furthermore in \cite{jia} Jiang L.S reduced the number of sources of risks from n to 1 at one try in pricing a basket option with the expiry  payoff  of  the  geometric  mean  of  n underlying  assets  without  any  use  of  change  of numeraire. 

The works \cite{jia, xug} make us confirm that there must be another general transformation (other than the change of numeraire) reducing the space dimension of multi-dimensional Black-Scholes partial differential equations and give us the clue of this article. 

From the results of \cite{jia, xug} about pricing options with foreign currency strike price (pricing the option in Dollars), we have the idea that the transformation of multiplication of two variables would reserve the form of Black-Scholes partial differential equations. This transformation can reduce the dimension of space variable, and furthermore, this transformation can be applied repeatedly if some condition holds, so in such a case we can reduce the number of sources of risks by two or more.  The result of Jiang \cite{jia} makes  us  to  find  more  general transformation.   

In this article we prove the invariance of the form of multi-dimensional Black-Scholes partial differential equations under the multiplicative transformation of variables and provide some examples where we can reduce the space dimension of multi-dimensional Black-Scholes partial differential equations with this transformation. 

According to our study, in pricing financial derivatives, the possibility to reduce dimension depends on the expiry payoff function structures.

The remainder of this article consists as follows. In section 2 we prove the invariance of the form of multi-dimensional Black-Scholes partial differential equations under the multiplicative transformation of variables. In section 3, we give such examples as

$\bullet$    Pricing options whose strike price is in a currency different from the stock price, 

$\bullet$    Pricing a basket option with the expiry payoff of the geometric mean, 

$\bullet$    Pricing a European call foreign currency option . 

In section 4, we consider a relationship between the invariance of the form of Black-Scholes partial differential equations and the invariance of the form of parabolic equations.

%
%

\section{The Invariance of the Form of Black-Scholes Equations}

Black-Scholes partial differential equations are one of main models in financial mathematics. The following partial differential equation
\begin{equation} \label{eq01}   
\frac{\partial V}{\partial t}+\frac{1}{2}\sum_{i,j=0}^n a_{ij}S_iS_j \frac{\partial^2 V}{\partial S_i \partial S_j} 
+ \sum_{i=0}^n (r-q_i)S_i \frac{\partial V}{\partial S_i}-rV=0,
\end{equation} 
is called the $(n+1)$-dimensional \textit{Black-Scholes equation} with risk free rate $r$.
Here $r > 0$ is risk free interest rate, $S_i$ the $i$-th underlying asset with dividend rate 
$q_i (i=1, \cdots, n)$, $A=\left[a_{ij}\right]_{i,j=0}^n$ a non-negative definite $(n+1)\times (n+1)$ matrix and $V(S_0, S_1,\cdots, S_n, t)$ the price of an option derived from underlying assets $S_0, S_1,\cdots, S_n$ at time $t$. The existence and representation of the solution to equation (1) are described in \cite{jia}.
The following theorem is our main result.\\

\noindent\textbf{Theorem 1}     
Under the transformation
\begin{equation} \label{eq02}
z_1=S_0^{\alpha_0}S_1^{\alpha_1}, ~~ z_i=S_i, ~~ i=2, \cdots, n,
\end{equation} 
equation \eqref{eq01} is transformed into an $n$-dimensional Black-Scholes equation with risk free rate $r$. 
In other words, the equation \eqref{eq01} has a solution of the form 
\begin{equation}\label{eq03}
V(S_0, S_1, \cdots, S_n, t) = U(z_1, \cdots, z_n, t).
\end{equation} 
where $U(z_1,\cdots, z_n, t)$  is a solution to the following $n$  dimensional Black-Scholes equation with risk free rate $r$:
 \begin{equation}  \label{eq04}
\frac{\partial U}{\partial t}+\frac{1}{2}\sum_{i,j=1}^n \bar{a}_{ij}z_iz_j \frac{\partial^2 U}{\partial z_i \partial z_j} 
+ \sum_{i=1}^n (r-\bar{q}_i)z_i \frac{\partial U}{\partial z_i}-rU=0.
\end{equation} 
Here $\bar{a}_{ij},~\bar{q}_i$ are provided as follows:
\begin{equation*}
\bar{a}_{ij} = \left\{ 
\begin{array}{ll}
a_{00}\alpha_0^2+a_{01}\alpha_0\alpha_1+a_{10}\alpha_1\alpha_0+a_{11}\alpha_1^2, & i=j=1,\\
a_{0j}\alpha_0+a_{1j}\alpha_1, & i=1, j=2, \cdots, n,\\
a_{i0}\alpha_0+a_{i1}\alpha_1, & i=2, \cdots, n, j=1,\\
a_{ij}, & i,j=2, \cdots, n,
\end{array} \right.
\end{equation*}
\begin{equation}  \label{eq05}
\bar{q}_k = \left\{ 
\begin{array}{ll}
r-\sum_{i=0}^1 \left( r-q_i-\frac{1}{2}a_{ii} \right) \alpha_i- \frac{1}{2}\sum_{i,j=0}^1a_{ij} \alpha_i \alpha_j, & k=1,\\
q_k, & k=2, \cdots, n.
\end{array} \right.
\end{equation}

\begin{proof} If we rewrite the derivatives of $V$ on $S_i (i = 0,\cdots, n)$ in \eqref{eq01} by the derivatives of $U$ on $z_i (i = 1,\cdots,n)$ using \eqref{eq03}, then we have
\begin{eqnarray*}
& & \frac{\partial V}{\partial S_0}S_0 = \alpha_0 \frac{\partial U}{\partial z_1}z_1, ~~ \frac{\partial V}{\partial S_1}S_1 = \alpha_1 \frac{\partial U}{\partial z_1}z_1, ~~\frac{\partial V}{\partial S_i}S_i = \frac{\partial U}{\partial z_i}z_i, ~~ i=2, \cdots, n,\\
& & \frac{\partial^2 V}{\partial S_0^2}S_0^2 = \alpha_0^2 \frac{\partial^2 U}{\partial z_1^2}z_1^2 + 
\alpha_0(\alpha_0-1) \frac{\partial U}{\partial z_1}z_1,\\
& & \frac{\partial^2 V}{\partial S_1^2}S_1^2 = \alpha_1^2 \frac{\partial^2 U}{\partial z_1^2}z_1^2 + 
\alpha_1(\alpha_1-1) \frac{\partial U}{\partial z_1}z_1,\\
& & \frac{\partial^2 V}{\partial S_0\partial S_1}S_0S_1 = \alpha_0\alpha_1 \frac{\partial^2 U}{\partial z_1^2}z_1^2 + 
\alpha_0\alpha_1 \frac{\partial U}{\partial z_1}z_1,\\
& & \frac{\partial^2 V}{\partial S_0\partial S_j}S_0S_j = \alpha_0 \frac{\partial^2 U}{\partial z_1\partial z_j}z_1z_j, ~~  j=2, \cdots, n\\
& & \frac{\partial^2 V}{\partial S_1\partial S_j}S_1S_j = \alpha_1 \frac{\partial^2 U}{\partial z_1\partial z_j}z_1z_j, ~~  j=2, \cdots, n\\
& & \frac{\partial^2 V}{\partial S_i\partial S_j}S_iS_j = \frac{\partial^2 U}{\partial z_i\partial z_j}z_iz_j, ~~  i,j = 2, \cdots, n.
\end{eqnarray*} 
If we expand the terms of second order derivatives in \eqref{eq01} as follows 
\begin{equation*}
\sum_{i,j=0}^n b_{ij} = \sum_{i,j=0}^1 b_{ij} + \sum_{i=0}^1\sum_{j=2}^n b_{ij} + \sum_{i=2}^n\sum_{j=0}^1 b_{ij} + \sum_{i,j=2}^n b_{ij},
\end{equation*} 
and substitute the above derivatives into here, then we can easily have \eqref{eq04} and \eqref{eq05}. Then the $n$-dimensional matrix $\bar{A}=\left[\bar{a}_{ij}\right]_{i,j=1}^n$ is evidently symmetric 
and non-negative. In fact, for any $\boldsymbol{\xi}=(\xi_1, \cdots, \xi_n)^\perp \in \mathbf{R}^n$ (the superscript $"\perp"$ denotes transpose), we have 
\begin{eqnarray*}
\boldsymbol{\xi}^{\perp}\bar{\mathbf{A}}\boldsymbol{\xi} & = & \sum_{i,j=1}^n \bar{a}_{ij}\xi_i\xi_j \\
& = & \bar{a}_{11}\xi_1^2+\sum_{j=2}^n \bar{a}_{1j}\xi_1\xi_j + \sum_{i=2}^n \bar{a}_{i1}\xi_i\xi_1 + 
\sum_{i,j=2}^n \bar{a}_{ij}\xi_i\xi_j \\
& = & \sum_{i,j=0}^1 a_{ij}\alpha_i\alpha_j\xi_1^2 + \sum_{j=2}^n \left( \sum_{i=0}^1 a_{ij} \alpha_i \right) \xi_1\xi_j +  \sum_{i=2}^n \left( \sum_{j=0}^1 a_{ij} \alpha_j \right) \xi_i\xi_1 + \sum_{i,j=2}^n a_{ij}\xi_i\xi_j \\
& = & \sum_{i,j=0}^1 a_{ij}(\alpha_i\xi_1)(\alpha_j\xi_1) + \sum_{i=0}^1\sum_{j=2}^n a_{ij} (\alpha_i\xi_1)\xi_j + \sum_{i=2}^n\sum_{j=0}^1 a_{ij}\xi_i (\alpha_j\xi_1) + \sum_{i,j=2}^n a_{ij}\xi_i\xi_j.
\end{eqnarray*} 
So if we let  $ \boldsymbol{\eta}=(\alpha_0\xi_1, \alpha_1\xi_1, \xi_2, \cdots, \xi_n)^\perp \in \mathbf{R}^{n+1}$, then from the nonnegativeness of $A$, we have
\begin{equation*}
\boldsymbol{\xi}^{\perp}\bar{A}\boldsymbol{\xi} = \boldsymbol{\eta}^{\perp}A\boldsymbol{\eta} \geq 0,
\end{equation*} 
which completes the proof of theorem 1. (QED)\\
\end{proof}

If we apply the transformation \eqref{eq02} repeatedly, then we can get the following corollary 1.
\begin{corollary}   
Let $k \in\{1, 2, \cdots, n\}$. Under the transformation 
\begin{equation}  \label{eq06}
z_k = S_0^{\alpha_0} \cdots S_k^{\alpha_k}, ~~ z_i = S_i, ~~ i=k+1, \cdots, n.
\end{equation} 
the equation \eqref{eq01} is transformed into an $(n+1-k)$-dimensional Black-Scholes equation. 
\end{corollary}

Let consider a terminal condition 
\begin{equation}  \label{eq07}
V(S_0, S_1, \cdots, S_n, T) = P(S_0, S_1, \cdots, S_n)
\end{equation} 
of $(n+1)$-dimensional Black-Scholes equation \eqref{eq01}. Then from theorem 1, we have the following corollary 2.

\begin{corollary}  
Assume that there exists an $n$-dimensional function $F$ such that   
\begin{equation}  \label{eq08}
P(S_0, S_1, S_2, \cdots, S_n) = F(S_0^{\alpha_0}, S_1^{\alpha_1}, S_2, \cdots, S_n).
\end{equation} 
Then by the change of variables given by \eqref{eq02} and \eqref{eq03}, the terminal value problem given by \eqref{eq01} and \eqref{eq07} is transformed into the $n$-dimensional Black-Scholes equation's terminal value problem given by \eqref{eq04} and 
\begin{equation} \label{eq09}
U(z_1, \cdots, z_n, T) = F(z_1, \cdots, z_n).
\end{equation} 
\end{corollary}

\textbf{Remark 1} As shown in \cite{hyo}, if the expiry payoff function $P(S_0, S_1, \cdots, S_n)$ has the homogeneity for its variables: 
\begin{equation*}
P(aS_0, aS_1, \cdots, aS_n) = aP(S_0, S_1, \cdots, S_n), \forall a>0,
\end{equation*} 
then by the change of numeraire 
\begin{equation} \label{eq10}
U=\frac{V}{S_0}, ~~ z_i=\frac{S_i}{S_0}, ~~ i=1, \cdots, n,
\end{equation} 
the $(n+1)$-dimensional terminal value problem given by \eqref{eq01} and \eqref{eq07} is  transformed into a terminal value problem for an $n$-dimensional Black-Scholes equation with risk free rate 0. The new transformed expiry payoff function $F$ is given by 
\begin{equation*}
F(z_1, \cdots, z_n) \triangleq P(1, z_1, \cdots, z_n).
\end{equation*} 
Then the new expiry payoff function $F$ no more has the homogeneity for its variables. For example   
\begin{equation*}
P(S_0, S_1, S_2) = \textnormal{max}(S_0, S_1, S_2)
\end{equation*} 
has the homogeneity but its new transformed 2 dimensional function   
\begin{equation*}
F(z_1, z_2) = \textnormal{max}(1, z_1, z_2)
\end{equation*} 
has no homogeneity. Thus the change of numeraire only can reduce the number of sources of risks by one and it is impossible to use repeatedly. \\

\textbf{Remark 2} In corollary 2, if the new expiry payoff $F$ has the same property with $P$ in (8), then we can use the transformation (2) repeatedly. For example, if   
\begin{equation*}
P(S_0, S_1, S_2) = \textnormal{max}(S_0S_1S_2-K, 0)
\end{equation*} 
and we apply the transformation $z = S_0\cdot S_1$, then the new expiry payoff $F$ is also given by 
\begin{equation*}
F(z_1, z_2) = \textnormal{max}(z_1z_2-K, 0),
\end{equation*} 
thus  it  is  also  the  function  of  $z_1\cdot z_2$  and  we  can  use  the  transformation \eqref{eq02}  once  more,  and  then  we  have  one-dimensional function $G(x) = \textnormal{max}(x-K, 0)$. \\

\textbf{Remark 3} In our opinion, in multi-dimensional Black-Scholes equations, the principle to reduce dimension is based on its invariance under the transformation  \eqref{eq06}  and  the  change  of 
 numeraire \eqref{eq10}.  Thus  in  pricing  financial  derivatives  problems  derived to multi-dimensional
Black-Scholes equations,  the  possibility  to  reduce  dimension  depends  on  the  structures  of  their  expiry  payoff functions. If the price model of a derivative can be derived to Black-Scholes equation and its expiry payoff is a function 
of  some  group  of  risk  source  variables  such  as  \eqref{eq06}  and  furthermore,  if  the  new  expiry  payoff  function  of  group variables has homogeneity, then using \eqref{eq06} and \eqref{eq10} we can reduce the dimension of the problem by two or more.   

%
%

\section{Examples}

%
%

\subsection{Options with Foreign Currency Strike Price }

This  example  is  studied  in  \cite{ben}  by  expectation  method  and  in \cite{hyo} by  PDE 
method. For details about the financial background, we refer to \cite{ben, hyo}.  Here we only mention the 
use of the transformation given by the equation \eqref{eq02}. 

\textbf{Problem:} The underlying stock is traded in UK pounds and the option exercise price is in US dollars. At $t = 0$ the option is an at-the-money option (that is, the strike price is the same with the underlying stock price \cite{jia, wil}) when the strike price is expressed in UK pounds. This pound strike price is converted into dollars at $t = 0$. The dollar strike price computed like this is kept constant during the life of the option. At the expiry date $t = T$, the option holder can pay the fixed dollar strike price to buy the underlying stock. Find the fair price of this option. 
 
\textbf{Mathematical model} 

Let denote by $S(t)$ the stock price (in UK pounds), $r_p$ the short rate in UK pound market, $r_d$ the short rate in US dollar, $X(t)$ dollar/pound exchange rate (then $Y(t) = X(t)^{-1}$ is pound /dollar exchange rate), $K_d$ the strike price expressed in US dollar and $K_p(t)$ the strike price expressed in UK pound.  
 
\textbf{Assumption}  1) The stock price $S(t)$ satisfies geometric Brown motion: 
\begin{equation*}
dS(t)=\alpha_SS(t)dt+\sigma_SS(t)dW^S(t)~ \textnormal{(objective measure)}.
\end{equation*}

\quad 2) $r_p$ and $r_d$ are deterministic constants. 

\quad 3) Dollar/pound exchange rate $X(t)$ satisfies Garman-Kohlhagen model \cite{gar}: 
\begin{equation*}
dX(t)=\alpha_XX(t)dt+\sigma_XX(t)dW^X(t)~ \textnormal{(objective measure)}.
\end{equation*}
Here $W^S(t)$ and $W^X(t)$ are the scalar Wiener processes and have the following relations:
\begin{equation*}
dW^S(t) \cdot dW^X(t) = \rho dt, ~~ |\rho|<1.
\end{equation*}

\textbf{Explanation about strike price}: $ K_p(0)=S(0),~ K_d = K_p(0) \cdot X(0) = S(0) \cdot X(0) \equiv \textnormal{constant}$. The  dollar  strike  price  is  constant  but  the  pound  strike  price  randomly  varies  as  a  result  of varying exchange rate: 
\begin{equation*}
K_p(t)=K_d \cdot X(t)^{-1} = S(0) \cdot X(0) \cdot X(t)^{-1}.
\end{equation*}

\textbf{Explanation about maturity payoff}: The US dollar price of the option at maturity is 
\begin{equation*}
F_d=\textnormal{max}(S(T)\cdot X(T)-K_d, 0).
\end{equation*}
Then the UK pound price of the option at maturity is 
\begin{eqnarray*}
F_p & = & \textnormal{max}(S(T)-K_p(T), 0) \\
	& = & \textnormal{max}(S(T)-S(0) \cdot Y(0)^{-1} \cdot Y(T), 0).
\end{eqnarray*}

Now we derive the PDE model for pricing the option in dollars. Let $V_d = V(S, X, t)$ be the price of the option in US dollars. By $\Delta$-hedging, construct a portfolio $\Pi$ as
\begin{equation*}
\Pi = V-\Delta_1 SX-\Delta_2X.
\end{equation*}   
(the dollar price of this portfolio consists of an option, $\Delta_1$  shares of stocks and $\Delta_2$ UK pounds.) 
Choose $\Delta_1, \Delta_2$  such that $\Pi$ is risk-free in $(t, t + dt)$, i.e. 
\begin{equation*}
d\Pi = r_d\Pi dt.
\end{equation*}   
Then we easily can derive the following PDE pricing model \cite{hyo}: 
\begin{eqnarray}
& & \frac{\partial V}{\partial t} +\frac{1}{2} \left[ \sigma_S^2 S^2 \frac{\partial^2 V}{\partial S^2}+ 2\rho \sigma_S \sigma_X SX \frac{\partial^2 V}{\partial S \partial X}+ \sigma_X^2 X^2 \frac{\partial^2 V}{\partial X^2} \right] \nonumber \\
& & \quad \quad +(r_p-\rho \sigma_S \sigma_X)S \frac{\partial V}{\partial S}+(r_d-r_p)X \frac{\partial V}{\partial X}-r_dV = 0,    \ \label{eq11} \\
& & V(S, X, T) = \textnormal{max}(S \cdot X-K_d, 0).   \label{eq12}
\end{eqnarray} 
Here we remind that $S$ is the variable representing stock price, $X$ the variable representing dollar/pound exchange rate, $T$ the maturity and $V$ the option price. 

The equation \eqref{eq11} is a two-dimensional Black-Scholes equation and $T$-payoff function 
\eqref{eq12} does not satisfy homogeneity, but it is a function of the group variable
\begin{equation} \label{eq13}
z=SX.
\end{equation}      
(This change of variable transforms the stock price expressed in UK pound into UD dollar price.)   
Thus from corollary 2 of our theorem 1, by the transformation \eqref{eq13} our problem is transformed into the following one dimensional problem:
\begin{eqnarray} \label{eq14}
& & \frac{\partial V}{\partial t} +\frac{1}{2} ( \sigma_S^2 + 2\rho \sigma_S \sigma_X +\sigma_X^2)z^2 \frac{\partial^2 V}{\partial z^2} + r_dz \frac{\partial V}{\partial z} - r_dV = 0, \nonumber \\
& & V(z, T) = \textnormal{max}(z-K_d, 0).
\end{eqnarray}  
The problem \eqref{eq14} can be seen as an ordinary UK dollar call option pricing problem. By the standard Black-Scholes formula,   
\begin{eqnarray*}
& & V(z, t)=zN(d_1)-K_de^{-r_d(T-t)}N(d_2),
\end{eqnarray*}
where
\begin{eqnarray*}
& & d_1 = \frac{\textnormal{ln}\frac{z}{K_d}+\left(r_d+\frac{1}{2}\sigma_{S, X}^2 \right)(T-t)}{\sigma_{S, X} \sqrt{T-t}}, \quad d_2 = d_1-\sigma_{S, X}\sqrt{T-t},  \\
& & \sigma_{S, X}^2 = \sigma_S^2+2\rho \sigma_S \sigma_X+\sigma_X^2.\nonumber
\end{eqnarray*}
If we return to the original variables $(S, X)$, then we have the dollar price of the option: 
\begin{eqnarray*}
& & V_d(S, X, t)=SXN(d_1)-K_de^{-r_d(T-t)}N(d_2),
\end{eqnarray*} 
where
\begin{eqnarray*}
& & d_1 = \frac{\textnormal{ln}\frac{SX}{K_d}+\left(r_d+\frac{1}{2}\sigma_{S, X}^2 \right)(T-t)}{\sigma_{S, X} \sqrt{T-t}}, ~~ d_2 = d_1-\sigma_{S, X}\sqrt{T-t}.\nonumber
\end{eqnarray*}      
Considering $V_p  = V_d \cdot X^{-1}$  and $K_d  = S(0)\cdot X(0)$, we have the pound price of option: 
\begin{eqnarray*}
& & V_p(S, X, t)=SN(d_1)-\frac{S_0X_0}{X}e^{-r_d(T-t)}N(d_2),
\end{eqnarray*} 
where
\begin{eqnarray*}
& & d_1 = \frac{\textnormal{ln}\frac{SX}{S_0X_0}+\left(r_d+\frac{1}{2}\sigma_{S, X}^2 \right)(T-t)}{\sigma_{S, X} \sqrt{T-t}},~~ d_2 = d_1-\sigma_{S, X}\sqrt{T-t}.\nonumber
\end{eqnarray*} 


\subsection{Basket Option with the Expiry Payoff of the Geometric Mean}

Assume that the prices $m$ underlying assets $S_0, \cdots, S_{m-1}$  follow geometric Brownian motions 
and the expiry payoff of an option is   
\begin{equation} \label{eq15}
V(S_0, \cdots, S_{m-1}, T)=(S_0^{\alpha_0}, \cdots, S_{m-1}^{\alpha_{m-1}}-K)^+
\end{equation}      
where $\sum \alpha_i = 1, \alpha_i \geq 0$.  Such  an  option  is  called  a  {\it basket  option}  with  the  expiry  payoff  of  the geometric mean of $m$ underlying assets \cite{jia}. 

The  mathematical  model  for  this  option  is  the  terminal  value  problem \eqref{eq01} and \eqref{eq15} of $m$-dimensional Black-Scholes equation when $n+1 = m$.   

Under the transformation   
\begin{equation}\label{eq16}
z=S_0^{\alpha_0}\cdot \cdots \cdot S_{m-1}^{\alpha_{m-1}}
\end{equation}     
we have 
\begin{eqnarray*}
& & S_i \frac{\partial V}{\partial S_i} = \alpha_i z \frac{\partial V}{\partial z},~~ i=0,\cdots,m-1,\\
& & S_i^2 \frac{\partial^2 V}{\partial S_i^2} = \alpha_i^2 z^2 \frac{\partial^2 V}{\partial z^2} + \alpha_i^2 z \frac{\partial V}{\partial z} - \alpha_i z \frac{\partial V}{\partial z}, \\
& & S_iS_j \frac{\partial^2 V}{\partial S_i\partial S_j} = \alpha_i\alpha_j z^2 \frac{\partial^2 V}{\partial z^2} + \alpha_i\alpha_j z \frac{\partial V}{\partial z}, ~~ (i \neq j). 
\end{eqnarray*} 
If we substitute the above derivative expressions into the equation (1) when $n+1=m$ and consider $\sum \alpha_i = 1$, then we have 
\begin{eqnarray*}
\frac{\partial V}{\partial t} &+& \frac{1}{2} \left[ \sum_{i,j=0}^{m-1}a_{ij} \alpha_i \alpha_j z^2 \frac{\partial^2 V}{\partial z^2} + \left( \sum_{i,j=0}^{m-1}a_{ij} \alpha_i \alpha_j - \sum_{i=0}^{m-1}a_{ii} \alpha_i \right)z \frac{\partial V}{\partial z} \right] \\
& + & \left( r- \sum_{i=0}^{m-1}q_i \alpha_i \right)z \frac{\partial V}{\partial z}-rV = 0 \\
\end{eqnarray*} 
Let denote  
\begin{eqnarray*}
& & \hat{\sigma}^2 := \sum_{i,j=0}^{m-1}a_{ij} \alpha_i \alpha_j, ~~
 \hat{q} :=\sum_{i=0}^{m-1}\left(q_i +\frac{1}{2}a_{ii} \right)\alpha_i - \frac{1}{2} \hat{\sigma}^2. 
\end{eqnarray*} 
Then  we  have  the following terminal value problem of one dimensional  Black-Scholes  equation  (standard  call option): 
\begin{eqnarray*}
& & \frac{\partial V}{\partial t} + \frac{1}{2} \hat{\sigma}^2z^2 \frac{\partial^2 V}{\partial z^2} + (r-\hat{q})z \frac{\partial V}{\partial z} -rV = 0, \nonumber \\
& & V(z, T)=(z-K)^+. 
\end{eqnarray*} 
If we use the standard Black-Scholes formula and return to the original variables, then we have the pricing formula of a {\it basket option} with the expiry payoff of the geometric mean: 
\begin{eqnarray*}
& & V(S_0, \cdots, S_{m-1}, t) = e^{-\hat{q}(T-t)}S_0^{\alpha_0} \cdot \cdots \cdot S_{m-1}^{\alpha_{m-1}}N(d_1)-Ke^{-r(T-t)}N(d_2),  
\end{eqnarray*} 
where
\begin{eqnarray*}
& & d_1 = \frac{\textnormal{ln}\frac{S_0^{\alpha_0} \cdot \cdots \cdot S_{m-1}^{\alpha_{m-1}}}{K} + \left(r-\hat{q}+\frac{1}{2}\hat{\sigma}^2 \right)(T-t)}{\hat{\sigma}\sqrt{T-t}}, ~~ d_2 = d_1-\hat{\sigma}\sqrt{T-t}. \nonumber 
\end{eqnarray*} 


\subsection{Pricing a European Call Foreign Currency Option}

This problem was studied in \cite{xug} under a general condition with stochastic short 
rates  and  stochastic  exchange  rates.  They  consider the option as  a  interest  rate derivative,  so  that the partial  differential  equations  of  the  prices  that  they  derived  are  not  standard  multidimensional Black-Scholes equations. Thus our result can not be directly applied to their equation. Fortunately, as  mentioned  in \cite{hyo},  the  prices  of  zero  coupon  bonds  under  the  Vasicek  model,  Ho-Lee model or Hull-White model follow geometric Brown motions and the corresponding short rate is a deterministic  function  of  the  price  of  zero  coupon  bond.  So  here  we  consider  the  option  as  a  zero coupon bond derivative, just as in \cite{hyo}, and then we apply our theorem. \\

\textbf{Problem:} A holder of an European call foreign currency option has a right to fix the exchange rate as a strike exchange rate. Find the fair price of this option.  

We denote by $r_1(t)$ the short rate in domestic currency, $r_2(t)$ the short rate in foreign currency,  $F(t)$ domestic currency / foreign currency exchange rate, $K$ the strike exchange rate (domestic / foreign), and $T$ the expiry date. Under this notation, the expiry pay off of our European call foreign currency option is given by 
\begin{equation*}
[F(T) - K]^+.\\
\end{equation*}   

\textbf{Assumptions:} All discussion is done under the risk neutral measure $Q$ (domestic martingale measure). In what follows, $a_1, a_2, b_1, b_2$  are all positive constants, $\sigma_1, \sigma_2, \sigma_3$ are linear independent constant vectors and $\left\{ W_t; 0 \leq t \leq T \right\} = \left\{ \left(W_t^1, W_t^2, W_t^3 \right); 0 \leq t \leq T \right\}$ is a standard 3 dimensional Wiener process satisfying the following conditions: 
\begin{equation*}
E(dW_t^i)=0, ~~ Var(dW_t^i)=dt, ~~ Cov(dW_t^i, dW_t^j) = 0 (i \neq j), 1 \leq i, j \leq 3.
\end{equation*}  

1) The domestic and foreign short rates follow Vasicek model: 
\begin{equation*}
dr_i = (b_i-a_ir_i)dt + \sigma_i \cdot dW(t), ~~ i=1, 2.
\end{equation*}  

2) The exchange rate $F(t)$ follows Garman-Kohlhagen model \cite{gar}:   
\begin{equation*}
dF(t) = F(t)(r_1(t)-r_2(t))dt + F(t) \sigma_3 \cdot dW(t).
\end{equation*}  

3) The price $V$ of the option in domestic currency is given as a deterministic function $V = C(r_1, r_2, F, t)$ of domestic short rate, foreign short rate and exchange rate and assume   
\begin{equation*}
V \in C^{2, 1} (D \times [0, T)), \quad D = (-\infty, \infty) \times  (-\infty, \infty) \times  (0, \infty).
\end{equation*}  

\textbf{Dynamics of the price of zero coupon bond:} Let denote the price of domestic zero coupon bond with maturity $T$ by $p_1(t, r_1; T)$ (in domestic currency)  and  denote  the  price  of  foreign zero  coupon  bond  with  maturity $T$  by  $p_2(t, r_2;  T)$  (in foreign currency).   Then the price of zero coupon bond $p_i(t, r_i; T)$ satisfies the following equation \cite{wil}:
\begin{eqnarray*}
& & \frac{\partial p_i}{\partial t} + \frac{1}{2} |\sigma_i|^2 \frac{\partial^2 p_i}{\partial r_i^2} + (b_i-a_ir_i-\lambda_i |\sigma_i|) \frac{\partial p_i}{\partial r_i} - r_ip_i = 0, \\
& & p_i(T, r_i; T)=1.\nonumber 
\end{eqnarray*} 
Here $\lambda_1  = 0$ is the price of domestic market risk (under the domestic martingale measure) and $\lambda_2 \neq 0$ is the price of foreign market risk (under the domestic martingale measure), $|\sigma|$ denotes the 
length of vector $\sigma$. And its solution is expressed by   
\begin{equation*}
p_i(t, r_i; T) = A_i(t, T)e^{-B_i(t, T)r_i}, ~ \frac{\partial p_i}{\partial r_i} = -B_i(t, T)p_i, ~ B_i(t, T)=\frac{1}{a_i}\left(1-e^{-a_i(T-t)}\right),
\end{equation*}  
and so the short rate $r_i(t)$ is a deterministic function of $p_i  = p_i(t, r_i, T)$: 
\begin{equation}  \label{eq17}
r_i = r_i(t, p_i) = -\frac{1}{B_i(t)} (\textnormal{ln}p_i-A_i(t)) = -\frac{1}{B_i(t)} \textnormal{ln}p_i + \frac{A_i(t)}{B_i(t)}.
\end{equation}
   
As  shown  in  \cite{hyo},  the  dynamics  of  the  zero  coupon  bond  price  follows  geometric 
Brown motion: In fact 
\begin{eqnarray*}
dp_i & = & \left( \frac{\partial p_i}{\partial t} + \frac{1}{2} |\sigma_i|^2 \frac{\partial^2 p_i}{\partial r_i^2} \right)dt + \frac{\partial p_i}{\partial r_i} dr_i \\
& = & \left( \frac{\partial p_i}{\partial t} + \frac{1}{2} |\sigma_i|^2 \frac{\partial^2 p_i}{\partial r_i^2} +(b_i-a_ir_i) \frac{\partial p_i}{\partial r_i} \right)dt + \frac{\partial p_i}{\partial r_i} \sigma_i \cdot dW(t) \\
& = & \left( rp_i + \lambda_i |\sigma_i| \frac{\partial p_i}{\partial r_i} \right)dt + \frac{\partial p_i}{\partial r_i} \sigma_i \cdot dW(t) \\
& = & \left( r_i - \lambda_i |\sigma_i|B_i(t) \right) p_idt + p_iB_i(t) \sigma_i \cdot dW(t).
\end{eqnarray*} 
Thus we get
\begin{eqnarray} \label{eq18}
& & dp_i = \alpha_i(t)p_idt + p_i \Sigma_i(t) \cdot dW(t), ~~ i=1, 2, \nonumber \\
& & \Sigma_i(t) = -B_i(t) \sigma_i. 
\end{eqnarray} 
 
\textbf{PDE Model and Solving:} Now we can attack the pricing problem. Denote by 
\begin{equation*}
p_i = p_i(t; T).
\end{equation*}
Since $p_i (T, T) = 1$, then the price of our option can be rewritten as   
\begin{equation} \label{eq19}
V_T =\textnormal{max}(Fp_2-Kp_1, 0).
\end{equation}
From the assumption 3) and the fact that $r_1(t) = r_1(t, p_1)$ and $r_2(t) = r_2(t, p_2)$, the domestic price of the option at time $t$ can be rewritten as a function  
\begin{equation*}
V = V(p_1, p_2, F, t) 
\end{equation*}
of the zero coupon bond prices. By $\Delta$-hedging, construct a portfolio $\Pi$ as  
\begin{equation*}
\Pi = V-\Delta_1p_1-\Delta_2p_2F-\Delta_3F. \quad \textnormal{(in domestic currency)}
\end{equation*}
This portfolio consists of an option, $\Delta_1$ shares of domestic zero coupon bond, 
$\Delta_2$ shares of foreign zero coupon bond and $\Delta_3$ units of foreign currency. 
Choose $\Delta_1, \Delta_2, \Delta_3$ such that $\Pi$ is risk-free in $(t, t + dt)$, i.e. $d\Pi = r_1 \Pi dt$.
This is equivalent to the following equality 
\begin{equation} \label{eq20}
dV - \Delta_1dp_1 - \Delta_2d(p_2F) -  \Delta_3dF - \Delta_3r_2dtF = r_1(t, p_1)(V-\Delta_1p_1-\Delta_2p_2F-\Delta_3F)dt.
\end{equation}
By the three dimensional It\^o formula, we have 
\begin{eqnarray*}
& & dV = \frac{\partial V}{\partial p_1}dp_1 + \frac{\partial V}{\partial p_2}dp_2 + \frac{\partial V}{\partial F}dF + \Big\{ \frac{\partial V}{\partial t} + \frac{1}{2} \Big[ \left \vert \Sigma_1 \right \vert ^2 p_1^2 \frac{\partial^2 V}{\partial p_1^2}  \\
& & \quad\quad\quad + \left \vert \Sigma_2 \right \vert ^2 p_2^2 \frac{\partial^2 V}{\partial p_2^2} + \vert \sigma_3 \vert ^2 F^2 \frac{\partial^2 V}{\partial F^2} +2\Sigma_1 \cdot \Sigma_2 p_1p_2 \frac{\partial^2 V}{\partial p_1\partial p_2} \\
& & \quad\quad\quad + 2\Sigma_1 \cdot \sigma_3 p_1F \frac{\partial^2 V}{\partial p_1\partial F} + 2\Sigma_2 \cdot \sigma_3 p_2F \frac{\partial^2 V}{\partial p_2\partial F} \Big] \Big\}dt,  \\
& & d(p_2F) = p_2dF + Fdp_2 + \Sigma_2 \cdot \sigma_3 p_2Fdt.
\end{eqnarray*} 
If we substitute above two expressions into \eqref{eq20}, then we get      
\begin{eqnarray*}
& & \left( \frac{\partial V}{\partial p_1}-\Delta_1 \right)dp_1 + \left( \frac{\partial V}{\partial p_2}-\Delta_2F \right)dp_2  + \left( \frac{\partial V}{\partial F}-\Delta_2p_2-\Delta_3\right)dF \\
& & \quad\quad\quad + \Big\{ \frac{\partial V}{\partial t} + \frac{1}{2} \Big[ \vert \Sigma_1 \vert^2 p_1^2 \frac{\partial^2 V}{\partial p_1^2} + \vert \Sigma_2 \vert^2 p_2^2 \frac{\partial^2 V}{\partial p_2^2} + \vert \sigma_3 \vert^2 F^2 \frac{\partial^2 V}{\partial F^2} \\
& & \quad\quad\quad + 2\Sigma_1 \cdot \Sigma_2 p_1p_2 \frac{\partial^2 V}{\partial p_1\partial p_2} + 2\Sigma_1 \cdot \sigma_3 p_1F \frac{\partial^2 V}{\partial p_1\partial F} + 2\Sigma_2 \cdot \sigma_3 p_2F \frac{\partial^2 V}{\partial p_2\partial F} \Big] \Big\}dt \\
& & \quad\quad\quad - \Delta_2 \Sigma_2 \cdot \sigma_3p_2Fdt - \Delta_3 r_2(t, p_2)dt \cdot F \\
& & \quad \quad= r_1(t, p_1)(V-\Delta_1p_1-\Delta_2p_2F-\Delta_3F)dt.
\end{eqnarray*} 
Here we choose $\Delta_1$ and $\Delta_2$ such that
\begin{equation*}
\frac{\partial V}{\partial p_1}-\Delta_1 = 0, ~~ \frac{\partial V}{\partial p_2}-\Delta_2F = 0, ~~ \frac{\partial V}{\partial F}-\Delta_2p_2- \Delta_3 = 0,
\end{equation*} 
equivalently,
\begin{equation*}
\Delta_1 = \frac{\partial V}{\partial p_1}, ~~ \Delta_2 = \frac{1}{F} \frac{\partial V}{\partial p_2}, ~~ \Delta_3 = \frac{\partial V}{\partial F}-\frac{p_2}{F} \frac{\partial V}{\partial p_2}.
\end{equation*} 
Then we have 
\begin{eqnarray} \label{eq21}
& & \frac{\partial V}{\partial t} + r_1(t, p_1) \frac{\partial V}{\partial p_1}p_1 + [r_2(t, p_2)-\Sigma_2(t) \cdot \sigma_3] \frac{\partial V}{\partial p_2}p_2 \nonumber\\
& & \quad + [r_1(t, p_1)-r_2(t, p_2)] F \frac{\partial V}{\partial F}+\frac{1}{2} \Big[ \vert \Sigma_1(t) \vert^2 p_1^2 \frac{\partial^2 V}{\partial p_1^2}  \nonumber \\
& & \quad + \vert \Sigma_2(t) \vert^2 p_2^2 \frac{\partial^2 V}{\partial p_2^2} + \vert \sigma_3 \vert^2 F^2 \frac{\partial^2 V}{\partial F^2} + 2\Sigma_1(t) \cdot \Sigma_2(t) p_1p_2 \frac{\partial^2 V}{\partial p_1\partial p_2} \\
& & \quad + 2\Sigma_1(t) \cdot \sigma_3 p_1F \frac{\partial^2 V}{\partial p_1\partial F} + 2\Sigma_2(t) \cdot \sigma_3 p_2F \frac{\partial^2 V}{\partial p_2\partial F} \Big]-r_1(t, p_1)V = 0.\nonumber
\end{eqnarray} 
The problem \eqref{eq21} and \eqref{eq19} is the pricing model for European call foreign currency option which is considered as a derivative of two country’s zero coupon bonds. 

The equation \eqref{eq21} has a simillar form of Black-Scholes equation  but  the  terms  of  first  order  derivatives  and  itself  of  unknown  function  have strongly varying coefficients which depend on space variables.  Although our theorem deals with constant coefficient Black-Scholes equation, but the change of variables 
\begin{equation*}
\boldsymbol{z = p_2 \cdot F}
\end{equation*} 
does work well. This change of variables composes the price of foreign zero coupon bond and the exchange rate to the  domestic  price  of  foreign  zero  coupon  bond.  By  this  change  of  variables,  the  space 3-dimensional problem given by \eqref{eq21} and \eqref{eq19} is transformed into the following space 2-dimensional problem: 
\begin{eqnarray} \label{eq22}
& & \frac{\partial V}{\partial t} + r_1(t, p_1) \frac{\partial V}{\partial p_1}p_1 +  r_1(t, p_1) \frac{\partial V}{\partial z}z + 
\frac{1}{2} \Big[ \vert \Sigma_1(t) \vert^2 p_1^2 \frac{\partial^2 V}{\partial p_1^2}+ \vert \Sigma_2(t) + \sigma_3 \vert^2 z^2   \nonumber \\
& & \quad \quad\quad \frac{\partial^2 V}{\partial z^2}+ 2\Sigma_1(t) \cdot (\Sigma_2(t)+\sigma_3)p_1z \frac{\partial^2 V}{\partial p_1\partial z} \Big]-r_1(t, p_1)V = 0, \\
& & V_T=\textnormal{max}(z-Kp_1, 0). \nonumber
\end{eqnarray} 
The original expiry payoff function \eqref{eq19} has no homogeneity on its variables $(p_1, p_2, F)$ but the changed expiry payoff function of the problem \eqref{eq22} has homogeneity on its new variables $(z, p_1)$, and thus by theorem 1 of [4] we can use the standard change of numeraire 
\begin{equation*}
U=\frac{V}{p_1}, ~~ y=\frac{z}{p_1} (=\frac{p_2F}{p_1}).
\end{equation*}
This change of variables transforms the bond price and domestic price of foreign zero coupon 
bond  into  relative  price  with  respect  to  the  zero  coupon  bond  price and we have the following terminal value problem of 1-dimensional Black-Scholes equation with risk free rate 0: 
\begin{eqnarray} \label{eq23}
& & \frac{\partial U}{\partial t} + \frac{1}{2} \vert \Sigma_1(t) - \Sigma_2(t) - \sigma_3 \vert^2 \frac{\partial^2 U}{\partial y^2}y^2 = 0, \\
& & U(y, T) = \textnormal{max}(y-K, 0). \nonumber
\end{eqnarray} 
We can easily solve \eqref{eq23} using standard method of [3]. The solution of  \eqref{eq23} is 
\begin{eqnarray*}
U(y, t) = yN(\bar{d}_1)-KN(\bar{d}_2). 
\end{eqnarray*} 
Here
\begin{eqnarray*}
& & \bar{d}_1 = \frac{\textnormal{ln}\frac{y}{K} + \frac{1}{2} \sigma^2(t, T)}{\sigma(t, T)}, \quad \bar{d}_2 =\bar{d}_1 -\sigma(t, T), \\
& & \sigma^2(t, T) = \int_t^T \vert \Sigma_1(u)-\Sigma_2(u)-\sigma_3 \vert^2du.
\end{eqnarray*} 
Considering  \eqref{eq18}, then we have  
\begin{equation*}
\sigma^2(t, T) = \int_t^T \vert B_1(u, T)\sigma_1 - B_2(u, T)\sigma_2 + \sigma_3 \vert^2du. 
\end{equation*}
Return to the original variables $V, p_1(t, T), p_2(t, T), F$, then we have the {\it price of European call 
foreign currency option}: 
\begin{equation}  \label{eq24}
V(p_1, p_2, F, t) = p_2(t, r_2, T)FN(d_1)-Kp_1(t, r_1, T)N(d_2),
\end{equation}
where
\begin{equation*}
d_1 = \frac{\textnormal{ln}\frac{p_2(t, r_2(t), T) \cdot F(t)}{p_1(t, r_1(t), T) \cdot K} + \frac{1}{2} \sigma^2(t, T)}{\sigma(t, T)},  \quad d_2 =d_1 -\sigma(t, T).
\end{equation*}

Note: The formula  \eqref{eq24} coincides with the pricing formula in \cite{xug}. 

%
%

\section{The Invariance of the Form in Parabolic Equation}

In fact, the invariance of the form of Black-Scholes equations is based on the invariance of the 
form in parabolic equation under a change of variables with the linear combination of variables. Using the theorem 1 and the change of variable ${x_i}  = {\ln}{S_i}$, we can easily get a transformation under which the form of parabolic equation is not changed and the dimension is reduced. 

As shown in \cite{jia}, the change of variable ${x_i}  = {\ln}{S_i} (i = 0, 1, \cdots, n)$ transforms the equation  \eqref{eq01} into a parabolic equation and we have the diagram:   
\begin{eqnarray*}
& & (1) \underset{x=\textnormal{ln}S}{\longleftrightarrow} \frac{\partial V}{\partial t}+\frac{1}{2}\sum_{i,j=0}^n a_{ij} \frac{\partial^2 V}{\partial x_i \partial x_j} + \sum_{i=0}^n \left( r-q_i-\frac{a_{ii}}{2} \right) \frac{\partial V}{\partial x_i}-rV = 0 \\
& & ~ \updownarrow (2) \qquad \qquad\qquad\qquad \qquad\qquad\qquad \updownarrow Tx=y \\
& & (4) \underset{y=\textnormal{ln}z}{\longleftrightarrow} \frac{\partial V}{\partial t}+\frac{1}{2}\sum_{i,j=1}^n \bar{a}_{ij} \frac{\partial^2 V}{\partial y_i \partial y_j} + \sum_{i=1}^n \left( r-\bar{q}_i-\frac{\bar{a}_{ii}}{2} \right) \frac{\partial V}{\partial y_i}-rV = 0 
\end{eqnarray*}
where new change of variables $\mathbf{y}=T\mathbf{x}$ is given by 
\begin{eqnarray*}
& & y_1 = \alpha_0x_0 + \alpha_1x_1, \nonumber\\
& & y_i = x_i, ~~ i=2, \cdots i=2, \cdots, n. 
\end{eqnarray*}
This change of variables reserve the form of parabolic equation and reduce the number of 
space variables. 

%
%

\section{Conclusions}

Multi-dimensional Black-Scholes equations have the form invariance under the change of 
variables product \eqref{eq02} and its space dimension is reduced under the change of variables. 

In the pricing problems of financial derivatives described as a terminal value problem for 
multi-dimensional  Black-Scholes  equation,  if  its  expiry  payoff  has  the  combination  of  variables 
such as \eqref{eq02}, then the space dimension can be reduced. 

In some pricing problems of interest rate derivative that have three or more risk resources 
and  are  not  described  by  Black-Scholes  equations  (for  example, \cite{xug}),  the  space
dimension can be reduced by two or more; the main reason is that the pricing problem is described 
as a simillar form of multi-dimensional Black-Scholes equation when we consider the interest rate derivative as a risk free zero  coupon  bond  derivative  and  its  expiry  payoff  function  has  not  only  a  combination  of variables but also homogeneity on the new group variables.  

The  method  of  considering  interest  rate  derivatives  as  no  coupon  bond  derivatives  is  still 
effective  in  any  interest  rate  models  satisfied  the  assumptions  "(i) the volatility of short rate $r$    does not depend on $r$, (ii) the price of zero coupon bond follows geometric Brown motion 
(see \eqref{eq18}), (iii) short rate $r$ is a deterministic function of the price of zero coupon bond 
(see \eqref{eq17})".  For examples, Vasicek, Ho-Lee and Hull-White models satisfy all these properties. For counter examples,  HJM model satisfies (i) and (ii) but does not satisfy (iii); CIR model satisfies (ii)  and  (iii)  but  does  not  satisfy  (i),  so  the  resulting  equations \eqref{eq22} and \eqref{eq23} after change  of  variables have the coefficients $\Sigma_1, \Sigma_2$ that still depend on $r$ or $p_1$  and $p_2$.  

If  an  asset  $F$  that  depends  on  the  short  rate  satisfies  the  above  three  assumptions,  then  we  can consider interest rate derivatives as $F$ derivatives and use Black-Scholes equations.  For a counter example, if $r$ is stochastic and satisfies (i), $B$ is bank account, that is 
\begin{equation*}   
B(t)=\textnormal{exp}\int_0^t r(u)du,
\end{equation*}
then  $B$  satisfies  (ii)  but  it does  not  satisfy  (iii),  so  we  cannot  consider  interest  rate  derivatives  as $B$ derivatives.


\end{document}